\documentstyle[12pt,epsfig]{article}

\newlength{\dinwidth}
\newlength{\dinmargin}
\begin{document}
\def\bold#1{\setbox0=\hbox{$#1$}%
     \kern-.025em\copy0\kern-\wd0
     \kern.05em\copy0\kern-\wd0
     \kern-.025em\raise.0433em\box0 }
\def\slash#1{\setbox0=\hbox{$#1$}#1\hskip-\wd0\dimen0=5pt\advance
       \dimen0 by-\ht0\advance\dimen0 by\dp0\lower0.5\dimen0\hbox
         to\wd0{\hss\sl/\/\hss}}
\def\lq{\left [}
\def\rq{\right ]}
\def\LL{{\cal L}}
\def\VV{{\cal V}}
\def\AA{{\cal A}}
\def\BB{{\cal B}}
\def\MM{{\cal M}}
\def\ovl{\overline}
\newcommand{\be}{\begin{equation}}
\newcommand{\ee}{\end{equation}}
\newcommand{\bea}{\begin{eqnarray}}
\newcommand{\eea}{\end{eqnarray}}
\newcommand{\nn}{\nonumber}
\newcommand{\dd}{\displaystyle}
\newcommand{\bra}[1]{\left\langle #1 \right|}
\newcommand{\ket}[1]{\left| #1 \right\rangle}
\newcommand{\qq}{<0|{\bar q} q|0>}
\newcommand{\spur}[1]{\not\! #1 \,}
\thispagestyle{empty}
\vspace*{1cm}
\rightline{BARI-TH/97-267}
\rightline{CPT-S 551.0797}
\rightline{August 1997}

\vspace*{3cm}

\begin{center}

{\LARGE \bf
Electromagnetic Mass Difference\\ of Heavy Mesons\\}
\vspace*{1cm}
{\large P. Colangelo$^{a}$, M. Ladisa$^{a,b}$, G. Nardulli$^{a,b}$,
T. N. Pham$^c$\\}
\vspace*{1.cm}

$^{a}$ Istituto Nazionale di Fisica Nucleare, Sezione di Bari, Italy\\
$^{b}$ Dipartimento di Fisica, Universit\'a di Bari, Italy\\
$^{c}$ Centre de Physique Th\'eorique, \\ 
Centre National de la Recherche Scientifique, UPR A0014, \\  
Ecole Polytechnique, 91128 Palaiseau Cedex, France

\end{center}

\vspace*{1cm}
\begin{abstract}
Using the Cottingham formula,
we give an estimate of the electromagnetic mass splitting of pseudoscalar
heavy mesons in the beauty and charm sector.
We include in the dispersion relation the Born term, the $1^-$
resonance and the positive parity $1^+$ resonance. We also
evaluate the contribution to the mass difference  from the 
isospin breaking quark mass differences. 
Our results: 
$m_{B^+}-m_{B^0} = - 0.83\pm 0.34 \,\rm MeV$ and
$m_{D^+}-m_{D^0} = + 4.33\pm 0.37 \,\rm MeV$, 
are in agreement with the experimental measurements:
$m_{B^+}-m_{B^0} = - 0.35\pm 0.29 \,\rm MeV$ and
$m_{D^+}-m_{D^0} = + 4.78\pm 0.10 \,\rm MeV$.
We also compute the mass differences in 
the infinite heavy quark mass limit, which  show small deviations from
the finite mass results for the $B$ case and $30\%$ effects in the 
charm case.
\noindent
\end{abstract}

\newpage
\baselineskip=18pt
\setcounter{page}{1}
                               
\section{Introduction}

The mass difference between $B^\pm$ and $B^0$ mesons 
is an interesting physical quantity,
whose precise knowledge might be of primary importance at the future
B-factories. As a matter of fact, the ratio 
$\displaystyle \frac{BR(\Upsilon(4S)\to B^0 {\bar B^0})} 
{BR (\Upsilon(4S)\to B^{+} {\bar B^-})}$
determines the relative abundance of neutral and charged $B$ mesons
produced at such accelerators and
is strongly dependent on the $B^{+}-B^{0}$ mass difference, since the 
$B$ pair production threshold is very close to the $\Upsilon(4S)$ mass. 

The experimental determination of
$\delta m (B^{+}-B^{0})$ changed significantly during the last
ten years, from the value 
$\delta m = -2.0\pm 1.1\pm 0.3\,\rm MeV$ \cite{cleo87} 
to the values 
$\delta m = +0.9\pm 1.2\pm 0.5\,\rm MeV$ (ARGUS) \cite{argus}
and 
$\delta m = +0.4\pm 0.6\pm 0.5\,\rm MeV$ (CLEO) \cite{cleo1}. 
A recent measurement 
by the CLEO Collaboration \cite{cleo2} gives a negative mass difference
$\delta m =-0.41\pm 0.25\pm 0.19 \rm \;MeV$, and  the combined
CLEO-ARGUS result is 
$\delta m = - 0.35 \pm 0.29 \,\rm  MeV$ \cite{PDG}.
Such a small value has to be compared to the analogous figure for
the $D^{+}-D^{0}$ mass difference:
$\delta m(D^{+}-D^{0})=+4.78\pm 0.10\,\rm  MeV$.

In the theoretical understanding of these values an important
role is played by the isospin symmetry breaking effects related to the
current quark masses. According to the modern  picture, which 
incorporates the old tadpole mechanism of Coleman and Glashow \cite{coleman}, 
the strong
isospin breaking is due to the intrinsic $u-d$ mass difference. By making
the $d$ quark heavier than the $u$ quark, one can explain,
at least qualitatively, all the known
meson and baryon electromagnetic mass differences \cite{isgur}. 
In particular,  the large value for the $D^{+}-D^{0}$ mass difference
can be explained by the combined effects of the $u-d$ mass
difference and the repulsive Coulomb energy between the $c$ and $\bar{d}$ 
quark. 

According to heavy quark symmetry, 
the effect due to the $u-d$ mass difference (qm)
is independent of the heavy quark mass; therefore,
in the case of  $B^{+}-B^{0}$ 
the quark mass term gives a large negative contribution and would
cancel out the repulsive Coulomb electrostatic energy resulting in a small 
$B^{+}-B^{0}$ mass difference.  Such a simple picture, however,
has to be implemented quantitatively, and this is the aim of the 
present letter. We begin by giving in Section 2 an estimate of the
contribution to $\delta m$ arising from the $u-d$ mass difference,
 using $SU(3)$ flavour symmetry and data on $B_s-B$ mass differences.
Since the tiny $B^{+}-B^{0}$ mass difference 
arises from the sum of two terms,
comparable in size (a few MeV), but opposite in sign, it is desirable 
to have an 
estimate of the electromagnetic contribution as accurate as possible.
This task is afforded by using 
the covariant Cottingham formula \cite{cottingham}, a method 
employed for the calculation of the electromagnetic mass differences
in the light hadron sector
\footnote{Previous
attempts to estimate the electromagnetic contribution to $\delta m(B^+-B^0)$
used the quark model \cite{isgur} and QCD sum rules \cite{kisslinger};
in \cite{isgur} 
the result  $\delta m(B^+-B^0)=-1.5 \,\rm MeV$, 
including quark mass effects, was obtained.
This issue has been investigated using the Cottingham formula in
\cite{goity,sundrum}. In \cite{goity} the elastic constribution is considered 
with a different treatment of the light quark currents, diregarding inelastic 
contributions. In \cite{sundrum} the calculations are carried out in the 
$N_c \to \infty$ limit. The numerical results are in agreement with the 
results of this paper.}.
By the Cottingham approach one relates the
electromagnetic (e.m.) mass difference to the forward Compton scattering
amplitudes $T_{1}$ and $T_{2}$, which  satisfy dispersion relations (DR)
and can be  put in a form which contains
 integration over space-like photon momenta 
$q^{2}=-Q^{2}<0$. 

The application of the Cottingham formula to the
evaluation
of electromagnetic mass differences has a long story 
 \cite{harari}. 
Previous (prior to QCD) attempts to use the Cottingham formula for
evaluating electromagnetic hadron mass differences encountered two problems: the
 first one is the convergence of the $Q^{2}$ integral and the second one is
 the convergence of the DR satisfied by $T_{i}$. The current approach to
 these problems involves a cut-off of the $Q^{2}$ integral 
 at a maximum value $Q^{2}_{max}=\mu^{2}$, where $\mu$ represents a scale 
coinciding with the onset of the QCD scaling behaviour \cite{collins}:
this point is discussed in Section 3. As for the convergence
of the DR, the different contributions to ${\rm Im}\,T_{i}$ can be related to
different Feynman graphs of an effective theory including hadrons and
photons. For light mesons this can be done by using chiral perturbation
theory, and recently some determinations of $\delta m(\pi^{+}-\pi^{0})$
and  $\delta m(K^{+}-K^{0})$ by chiral perturbation theory have appeared
in the literature \cite{donoghue}. 
In this approach, the subtraction constant can  be
computed directly from the Feynman amplitudes. A similar effective
theory was developed also for heavy mesons (for a review see 
\cite{casalbuoni}), and we will use it in our description of
the electromagnetic coupling of the heavy mesons involved in the
calculations
(the low-lying $B$ and $B^*$ mesons and the
first excited positive parity resonances). Therefore, also in our approach 
the subtraction constant in the DR is directly evaluated from 
the Feynman amplitude. These points are discussed in Section 4.

We conclude the paper by computing in Section 5 the meson mass differences
in the $m_{Q}\to \infty$ limit.
This calculation allows 
a remarkable simplification of the formalism, with a clear view of
the mechanism producing $\delta m$. We  find that the infinite
limit can be well applied
to the $B$ case, whereas in the charm case the deviation due to the finite 
heavy quark mass is of the order of $30\%$.
 
\section{Quark mass contributions}
The contribution  $\delta m(B^{+} - B^{0})_{\rm qm}$ and
$\delta m(D^{+} - D^{0})_{\rm qm}$  from the strong isospin breaking $u - d$
quark mass difference:
$m_{u} - m_{d}$, can be computed observing that
the (approximate) $SU(3)$ flavour symmetry allows us to write:
\be
\delta m^2(B^+-B^0)_{\rm qm}=~(m_u-m_d)<B^+|{\bar u} u|B^+>~
\simeq~(m_u-m_d)<B_s|{\bar s} s|B_s> \;\;\;.
\ee
 Considering the quark mass contribution to the $B_{s} - B^{0}$ and
$B_{s} - B^{+}$ mass differences, we have  similarly
\be
{\delta m}^{2}(B_s-B^+)_{\rm qm} +{\delta m}^{2}(B_s-B^0)_{\rm qm} =~2
\lq m_s -\frac{ m_u + m_d}{2}\rq <B_s|{\bar s} s|B_s>~.
\ee
For $\delta m(B_{s} -B)$ we can assume that the quark mass contribution
basically coincides with $\delta m$, since it is of the order of the
strange quark mass ($\simeq 100 MeV$), i.e. much larger than the
expected
electromagnetic mass difference (of the order 
 $\alpha \Lambda_{QCD} \simeq$ a few MeV). Writing
\be
{\delta m}^{2}(B_s-B)_{\rm qm}\simeq {\delta m}^{2}(B_{s}-B)
\ee
we obtain
\be
{\delta m}^2(B^+-B^0)_{\rm qm} =\lq {\delta m}^{2}(B_{s}-B^+) +
{\delta m}^{2}(B_{s}-B^0)\rq
\frac{m_u-m_d}{2 m_s - (m_{u}+m_{d})}~.
\ee
A similar formula holds for ${\delta m}^2(D^+-D^0)_{\rm qm}$. Using 
experimental data for ${\delta m}^{2}(B_{s}-B^{+})$, 
${\delta m}^{2}(B_{s}-B^{0})$, 
${\delta m}^{2}(D_{s}-D^{+})$, 
${\delta m}^{2}(D_{s}-D^{0})$ \cite{PDG}
and the result given in \cite{leut} for the (scale independent) ratios
of current quark masses from chiral perturbation theory:
\be
\frac{m_s-(m_u+m_d)/2 } {m_d-m_u}=40.8\pm 3.2~,
\ee we obtain:
\bea
{\delta m}(B^{+}{}-B^{0})_{\rm qm}&=&-2.23\pm 0.27\,\rm MeV
\label{res1} \\
{\delta m}(D^{+}{}-D^{0})_{\rm qm}&=&+2.54\pm 0.21\,\rm MeV \;\;\;.
\label{res2}
\eea
In the limit $m_{Q} \to \infty$ we expect
${\delta m}(B^{+}{}-B^{0})_{\rm qm}=-{\delta m}(D^{+}{}-D^{0})_{\rm qm}$
regardless of $m_{Q}$, a prediction which is supported by the results
(\ref{res1},\ref{res2}).

\section{The Cottingham formula}

Let us consider the mass splitting
of heavy mesons due to the electromagnetic interaction. To be definite,
we consider the $B$ meson; its
electromagnetic mass shift  can be derived by computing:
\be
\delta m^2 = \frac{ie^2}{2}  
\int\frac{d^4q}{(2\pi)^4} \frac{g_{\mu\nu}T^{\mu\nu}(q,p)}{q^2+i\epsilon}
\label{1}
\ee
where
\be
T^{\mu\nu}(q,p)=i\int d^4x e^{-iqx}<B(p)|T(J^\mu(x)J^\nu(0))|B(p)> \;\; ;
\label{2}
\ee
$J^\mu$ is the electromagnetic current. 
The Compton amplitude can be decomposed in terms of gauge 
invariant tensors:
\be
T^{\mu\nu}(q, p) = D_1^{\mu\nu}T_1(q^2,\nu) + D_2^{\mu\nu}T_2(q^2,\nu) 
 \label{3} 
\ee
($\nu=p\cdot q$), where
\bea
D_1^{\mu\nu} &=& -g^{\mu\nu} + \frac{q^\mu q^\nu}{q^2} \label{4} \\
D_2^{\mu\nu} &=& \frac{1}{m_B^2} \left( p^\mu-\frac{\nu}{q^2}q^\mu\right)
\left( p^\nu-\frac{\nu}{q^2}q^\nu \right) \;\;\;, \label{5}
\eea
$m_B$ being the meson mass.\par
The first step in the calculation of the integral (\ref{1}) 
consists of a rotation in the complex plane and a change of 
variables.
Let us consider the meson rest frame, $\nu = m_B q_0$. Since the singularities 
in $T^{\mu\nu}$ are located just below the positive real axis and just above 
the negative real axis in the complex $q_0$ plane,  
the integration over $q_0$ may be rotated to the imaginary axis 
$q_0=i\,k_0$ without encountering any singularity. After this transformation,
the integral involves only spacelike momenta for the photon, i.e.
$q^2=-Q^2=-(q_0^2+{\bf q}^2)$. After 
a change of variables from $(|{\bf q|},q_0)$ to $(Q^2, k_0)$, 
one obtains the Cottingham  formula~\cite{cottingham}:
\bea
\delta m^2 &=& \frac{e^2}{16\pi^3}\int_0^{\mu^2} \frac{dQ^2}{Q^2}
\int_{-\sqrt{Q^2}}^{+\sqrt{Q^2}}
dk_0  \sqrt{Q^2-k_0^2}~\times 
 \nn \\
 &\times &\left[ 
-3~T_1(-Q^2,i k_0)+ \Big(1-\frac{k_0^2}{Q^2}\Big)T_2(-Q^2,ik_0)\right] \;\;\;.
\label{6}
\eea

In eq.(\ref{6}) we have introduced a cut-off in 
the $Q^2$ integration at $Q^2_{max}=\mu^2$. Its origin is as follows 
(see \cite{collins} for a detailed discussion). To take into account 
possible ultraviolet (UV) divergences, the Cottingham formula has to be
renormalized. The renormalization is accomplished by a regularization of the
$Q^2$ integral and the inclusion of counterterms in the lagrangian describing
electromagnetic and strong interactions of quarks, gluons and photons. Both the
strong coupling constant $\alpha_s$ and the quark masses 
$m_q$ have to be specified at a renormalization 
mass scale $\mu$; since the counterterms cancel the infinite 
contribution induced by virtual particles with momenta larger than 
$\mu$, the net effect is analogous to a cut-off of the $Q^2$ integral at
$Q^2_{max}=\mu^2$. There is a residue smooth dependence on $\mu$, but it
should be canceled by the $\mu-$dependence of the renormalized 
quark masses and strong coupling constant. Typical values of $\mu$ are in 
the range of 1-2 GeV, corresponding to the onset of the scaling 
behaviour of QCD. The presence of heavy quarks does not change this procedure
since the relevant mass scale, in the infinite heavy quark mass limit,
is the residual energy release and the onset of scaling is again 
at a few GeV in this variable.

\par
The Compton amplitudes $T_{i}\, (i= 1,2)$  satisfy 
 dispersion relations (DR) in the $\nu = p\cdot q$ variable with 
$T_1$ requiring one subtraction \cite{cottingham}, as follows:
\bea
T_1(q^2,\nu)&=&T_1(q^2,0) + \frac{\nu^2}{\pi}\int_0^\infty
\frac{d\nu'^2}{\nu'^2}\frac{{\rm Im}\,T_1(q^2,\nu')}{\nu'^2 - \nu^2}\label{7}
\\
T_2(q^2,\nu)&=&\frac{1}{\pi}\int_0^\infty d\nu'^2\frac{{\rm Im}\,T_2(q^2,\nu')}
{\nu'^2 - \nu^2}\ \;\;\;. \label{8}
\eea
By employing these DR,
the integral over $k_0$ in eq.(\ref{6}) can be performed explicitly, 
with the result:
\bea
\delta m^2 &=& \frac{\alpha}{4\pi}
\int_0^{\mu^2} dQ^2 \Big[ -\frac{3}{2}~T_1(-Q^2,0) + 
 3\int_0^\infty\frac{d\nu'^2}{\nu'^2}W_1(-Q^2,\nu')
\Lambda_1(\frac{\nu'^2}{m_B^2 Q^2})\nn\\
&+ &\int_0^\infty \frac{d\nu'^2}{m_B^2 Q^2}W_2(-Q^2,\nu')
\Lambda_2(\frac{\nu'^2}{m_B^2Q^2})\Big] \;\;\;,\label{9}
\eea
where
\bea
\frac{1}{\pi}{\rm Im}\,T_i(q^2,\nu)&=&W_i(q^2,\nu)\label{10}
\eea
and
\bea
\Lambda_1(y)&=&\frac{1}{2} + y - y \sqrt{1+\frac{1}{y}} \label{l1}
\\
\Lambda_2(y)&=&-\frac{3}{2} - y + (1 + y)\sqrt{1+\frac{1}{y}}  \;\;\;.
\label{l2}
\eea 

\section{$B$ and $D$ meson electromagnetic mass differences}

In order to evaluate the DR (\ref{7}),(\ref{8}) we consider the contribution of
the Born term (the $B$ meson), the $J^P=1^-$ resonance $B^*$, 
and the positive parity resonance $J^P=1^+$ $B_1$. We 
notice that the Born term pole and the $B^*$ belong to the supermultiplet
$s_{\ell}^P=\left( \frac{1}{2} \right) ^-$ of the Heavy Quark Effective
Theory (HQET) ($s_{\ell}^P$ is the total angular momentum 
of the light degrees of freedom), whereas $B_1$ is the $J^P=1^+$ 
partner of the $s_{\ell}^P=\left( \frac{1}{2} \right)^+$ supermultiplet
(the other partner, with $J^P=0^+$, has no electromagnetic coupling to the 
$B$ meson). Let us observe explicitly that we do not introduce  the
 $s_{\ell}^P=\left( \frac{3}{2} \right)^+$ supermultiplet of 
heavy mesons containing the $J^P=1^+$ and $J^P=2^+$ states, for 
which we do not have  sufficient phenomenological information at the moment.

To compute the electromagnetic contribution to the $B$ meson mass difference, 
we consider the following matrix elements ($q=p'-p$):
\bea
<B(p')| J_{\rm em}^{\mu} |B(p)> &=& 
 f(q^2) (p + p')^\mu \label{13}\\
<B^*(p',\epsilon)| J_{\rm em}^{\mu} |B(p)> &=& 
i  h(q^2) 
\epsilon^{\mu\lambda\rho \sigma} \epsilon^*_\lambda q_\rho p_\sigma 
\label{14}\\
<B_1(p',\epsilon)| J_{\rm em}^\mu |B(p)> &=& 
-\frac{1}{2 m_B} \Big[ K_1[g^{\mu\sigma} (p^{\prime 2} - m^2_B)
-q^\sigma (p+p')^\mu]\nn \\
 & + & K_2 [q^2 g^{\mu\sigma}-q^\sigma q^\mu] \Big] 
~\epsilon^*_\sigma  \label{15} 
\eea
where $\epsilon$ is the $B^{*}$ or $B_1$ 
polarization vector
and  $f,~h,~K_1,~K_2$ are electromagnetic
form factors. In general they contain two 
terms, describing the couplings of the
electromagnetic current to the heavy $Q=b,c$ and 
light $q=u,d,s$ quarks, respectively:
\bea
f(q^2) &=& e_Q\xi(\omega)  + \frac{e_q}{1- q^2/m_V^2  }
\label{16}
\\
h(q^2) &=& \frac{e_Q}{\Lambda_Q} \xi(\omega)+
    \frac{e_q}{\Lambda_q (1- q^2/m_V^2)}  
\label{17}
\\
K_1(q^2) &=& 2 e_Q \tau_{1/2}(\omega)
+  \frac{e_q\sigma}{1- q^2/m_V^2}  \\
K_2(q^2)  &=& 2 e_Q \tau_{1/2}(\omega)
 \label{18}
\eea
where $\omega=v\cdot v^\prime$, and $v$ and $v^\prime$ are the heavy particle 
four velocities. We note explicitly that,
e.g. for
$B^+=u{\bar b}$, one has $\dd{e_q=\frac{2}{3}}$, $\dd{e_Q=e_{{\bar b}}}$= $
\dd{+\frac{1}{3}}$.
$\xi(\omega)$ is the Isgur-Wise form factor \cite{IW} and $\tau_{1/2}(\omega)$ 
is the analogous form factor describing the transitions
between the $(0^-,1^-)$ and the $(0^+,1^+)$  doublets of heavy mesons
\cite{iw2}.
From 
HQET \cite{IW}, at the leading order in $1/m_Q$:
\bea
<B(v^\prime)|{\bar b}\gamma^\mu b|B(v)>&=& m_B
( v^\mu+v^{\prime \mu})~ \xi(\omega)\label{19}\\
<B^*(v^\prime,\epsilon)|{\bar b}\gamma^\mu b|B(v)>&=&
i m_B \xi(\omega)\epsilon^{\mu\lambda\rho\sigma}~
\epsilon^{*}_\lambda v^\prime_\rho v_\sigma \label{20}\\
<B_1(v^\prime,\epsilon)|{\bar b}\gamma^\mu b|B(v)>&=&
2 \sqrt{m_B m_{B_1} }\tau_{1/2}(\omega) [(1-\omega)g^{\mu\sigma}+v^{\prime\mu}
v^\sigma ]\epsilon_\sigma^*~~. \label{21}
\eea

We can write $\xi(\omega)$ as 
\be
\xi(\omega)=\lq \frac{2}{1+\omega}\rq ^{2\rho^2}\;\;\;, 
\ee
using the normalization condition 
$\xi(1)=1$. The experimental determination of the slope 
$\rho^2$ contains several
uncertainties (see for example the discussion in \cite{mor}).
A value $\rho^2=1\pm 0.3$ encompasses
most of the theoretical predictions, while being in agreement with the data
\cite{rho}.
Therefore we shall take  in the following  $\rho^2=1$, which means that
we can take for the Isgur-Wise function the following expression
(with $\dd{\omega =1-\frac{q^2}{2 m^2_B}}$ in this case): 
\be
\xi(\omega)= \lq \frac{2}{1+\omega}\rq^2~=~\frac{1}{[1- q^2/4 m_B^2]^2}  
. \label{22} 
\ee
For $\tau_{1/2}(\omega)$ we take the 
QCD sum rule results
given in  \cite{paver}; we shall discuss
the uncertainties related to this choice below.
For the light quarks part of the 
electromagnetic current, we 
assume  Vector Meson ($\rho,~\omega$) Dominance of the form factor; under this
hypothesis the constants $\Lambda$ and $\sigma$ can be estimated
as follows: $\Lambda_Q = m_B$, $\Lambda_q 
\simeq 0.5 \,\rm GeV$ \cite{defazio}, $\dd{\sigma = 2\sqrt{2}~\frac{g_Vf_V}
{m_V^2}~|\mu| \sqrt{ m_B m_{B_1}}\simeq 2.7}$ 
(here $g_V\simeq 5.8$ , $f_V\simeq 0.17 \,\rm GeV^2$, 
$ m_V$ is the $\rho$ meson mass, and $|\mu|\simeq 0.1\, \rm GeV^{-1}$ 
parametrizes the $BB_1V$ vertex \cite{casalbuoni}.

\par
Using the matrix elements and the coupling constants just 
introduced, we can calculate the 
electromagnetic contribution to the mass splitting of  heavy mesons.
The contributions of the different terms to the DR are as
follows; the subtraction term  $T_1(q^2,0)$ is given by:
\bea
T_1(q^2,0) &=&
- 2 \left[f_+^2(q^2) - f_0^2(q^2)\right] +
 2 m_B^2 \left[h_+^2(q^2) - h_0^2(q^2)\right] 
 \nn  \\
&-&\frac{q^4}{4 m_B^2\nu_R} \left[(K_{1,+}+K_{2,+})^2 -(K_{1,0}+K_{2,0})^2 
\right]~,
\eea
where 
\be
\nu_R=\frac{q^2+m^2_B-m^2_{B_1}}{2}~.
\ee
As for the
two structure functions  $W_{1,2}(q^2,\nu)$ that appear
in the 
dispersion relations for $T_i$, they are given  by:
\bea
W_1(q^2,\nu) &=&- \frac{q^4}{2} \left[h_+^2 (q^2) - 
h_0^2(q^2)\right] \left[ \frac{q^2}{4}-m^2_B \right] 
\delta\left( \nu^2 - \frac{q^4}{4}
\right) \nn\\
&-&\frac{\nu_R}{4 m^2_B}\delta\left( \nu^2-\nu^2_R\right) {\times} \nn\\
&{\times}& \left\{  \left[q^2 (K_{1,+} + K_{2,+}) - 2 \nu_R K_{1,+} \right]^2
- \left[q^2 (K_{1,0} + K_{2,0}) - 2 \nu_R K_{1,0} \right]^2\right\}
\\
W_2(q^2,\nu) &=&
- 2 
m_B^2 q^2 \left[f_+^2(q^2)-f_0^2(q^2)\right]\delta \left( \nu^2-\frac{q^4}{4}
\right) \nn\\
& +& \frac{q^4 m_B^2 }{2}  \left[h_+^2(q^2) - 
h_0^2(q^2)\right]  \delta \left(\nu^2 -\frac{q^4}{4} \right)+ \nn     \\
&-&\frac{\nu_R q^2 }{4 m^2_{B_1}} \delta \left( \nu^2-\nu^2_R\right){\times}
 \nn \\
&{\times}& \left\{  \left[q^2 (K_{1,+} - K_{2,+})^2 - 4 m_{B_1}^2 K^2_{1,+}
\right] 
- \left[q^2 (K_{1,0} - K_{2,0})^2 - 4 m_{B_1}^2 K^2_{1,0}\right] \right\}
\eea
where $h_+,~f_+,~K_{j,+}$ refer to $B^+$ (resp. $D^+$) and 
and $h_0,~f_0,~K_{j,0}$ to $B^0$ (resp. $D^0$). 

The $1^-$ resonance is quite narrow (less than 1 keV); on the contrary,
the $1^+$ axial vector resonance
is broad enough to require the convolution of the mass 
difference term, depending upon ${m}_{{B}_{1}}$, 
with a lorenztian distribution centered on ${m}_{{B}_{1},\rm aver} 
=5.732$ GeV. Experimental data suggest a width $\Gamma=145$\, MeV.
The $D$ case is computed in full analogy with the $B$ one.
The numerical results of this analysis are reported in Table I,
for $\mu=1 \, \rm GeV$ and (in parentheses) $\mu=2\,\rm GeV$. 
It may be useful to stress
that the results are remarkably insensitive to variations of
the cut-off $\mu$ ; for example varying 
$\mu$ in the range $\mu=2-5 $ GeV introduces an uncertainty of less
than $8\%$. Another possible source of error is in the slope
of the Isgur-Wise function $\rho^2$. We find an uncertainty of $\pm 1\%$
for ${\delta m}(D^+-D^0)$ and negligible for ${\delta m}(B^+-B^0)$
when $\rho^2$ varies between $0.80$ and $1.20$. Also the uncertainties 
related to the choice of $\tau_{1/2}$ are negligible, given the smallness of
the $1^+$ contribution, and we do not expect significant contributions from
the $s_{\ell}=\left(\frac{3}{2}\right)^+$ poles.
\par
To compare our result to the experimental data we have to add the quark
mass
contribution computed in Section 2; the different terms and the total
theoretical prediction are reported in Table II which
shows a good agreement with experiment within the errors.

\section{Electromagnetic mass difference in the $m_Q\to \infty$ limit}

We wish now to evaluate the electromagnetic mass difference $B^+-B^0$
in the infinite heavy quark mass limit, which allows a remarkable simplification
of the formulae and a deeper understanding on the underlying physics.

In the $m_b \to \infty$ limit, since 
\begin{equation}
\delta m^2(B^+-B^0)=  2 m_B \,\delta m(B^+-B^0)~, \label{3.1}
\end{equation}
we get, in the $m_b \to \infty$ limit,
from previous formula, the result:
\begin{equation}
\delta m(B^+-B^0)_{\rm em}\to \delta m_{\rm Born}+\delta m_{V}+
 \delta m_{\rm subtr}
~~~~(m_b\to \infty)~, \label{3.2}
\end{equation}
where $\delta m_{\rm Born} $ is the contribution from the Born term and
is given by
\begin{equation} 
\delta m_{\rm Born}=\frac{\alpha m_V}{6 \pi}\left[ 5 \,\arctan\frac{\mu}{m_V}+
\frac{\mu/m_V} {1+\mu^{2}/m_V^{2}}\right]~. \label{3.3}
\end{equation}
As explained above, $m_V\simeq 770 \,\rm MeV$ is the $\rho$
 mass and $\mu$ is the 
cut-off; for $\mu=1\,\rm GeV$ and 
$\mu=2\,\rm GeV$, we get $\delta m_{\rm Born}= ~1.5\,\rm MeV$ 
and $1.9 \,\rm MeV $ respectively.
The remaining contributions in
(\ref{3.2}) arise from the subtraction term 
$T_1(-Q^2,0):~ \delta m_{\rm subtr}$,
and from the vector meson $1^{-}$ dispersive 
contribution to $W_1$ and $W_2$.
$\delta m_V$ (the contribution from the $1^{+}$ pole vanishes). The two terms 
are:
\begin{equation}
\delta m_{\rm subtr}=-\frac{\alpha \mu^2}{8 \pi \Lambda_q^2}\frac{m_B}{
1+\mu^2/m_V^2} \label{3.4}
\end{equation}
\begin{equation}
\delta m_{V}=\frac{\alpha \mu^2}{8 \pi \Lambda_q^2}\frac{m_B}
{1+\mu^2/m_V^2}-\frac{\alpha m^3_V}{12 \pi \Lambda_q^2}\left[
\arctan \frac{\mu}{m_V}- \frac{\mu/m_V}{1+\mu^{2}/m_V^{2}
}\right] \label{3.5}
\end{equation}
$(\Lambda_q\simeq 500$ MeV is the hadronic scale defined by eq. (\ref{17})). It
is interesting to observe that, while individually the subtraction
contribution (\ref{3.4}) and the vector meson contribution (\ref{3.5})
diverge in the infinite heavy quark mass limit, their sum is finite.
Therefore, summing up the three terms ($m_b\to\infty$), 
we obtain:

\begin{equation}
\delta m(B^+-B^0)_{\rm em}=\frac{\alpha m_V}{6 \pi }\left[ 
\left(5 - \frac{m_V^2}{2 \Lambda_q^2}\right)
\arctan \frac{\mu}{m_V}+\left(1 + \frac{m_{V}^{2}}{2\Lambda_{q}^{2}}\right)
\frac{ \mu/m_V } {{1+\mu^2/m_V^2}}\right] \;\;\;.\label{3.6}
\end{equation}

This gives, at $\mu=1$ GeV and  $\mu=2$ GeV,
$\delta m(B^+-B^0)_{\rm em}=1.36 $ MeV  and  $1.59 \, \rm MeV$ respectively,
which is remarkably
 close to the value obtained at finite mass and reported in Table II.
 A similar formula holds for $\delta m(D^+-D^0)_{\rm em}$ 
in the same limit ($m_c\to \infty$):
\begin{equation}
\delta m(D^+-D^0)_{\rm em}=\frac{\alpha m_V}{6 \pi }\left[\left(7+ 
\frac{m_V^2}{2 \Lambda_q^2}\right) \arctan\frac{\mu}{m_V}-\left(1 + 
 \frac{m_{V}^{2}}{2\Lambda_{q}^{2}}\right)
\frac{\mu/m_V}{1+\mu^2/m_V^2}\right] \label{3.7}
\end{equation}
where the differences with (\ref{3.6}) are only due to quark charge
factors. Numerically we find, at $\mu=1\,\rm GeV$ : 
$\delta m (D^+-D^0)_{\rm em}=1.92 $ MeV (this value is $2.72 \,\rm MeV$
for $\mu=2\,\rm GeV$). These results, valid
for $m_c\to\infty$, show significant deviations from the finite mass result 
 reported in Table I.

Besides showing the exact cancellation
of the divergent term in (\ref{3.4}) and (\ref{3.5}), which
confirms the scaling law $\delta m \to {\rm const}~~(m_b\to\infty)$, 
the previous analysis is interesting also because it explicitly shows 
the small dependence of the $m_b\to\infty$ results 
on the renormalization scale $\mu$.

 We also remark that, although the $1^{-}$ state makes only a small 
contribution to the  electromagnetic mass difference, its contribution
seems to increase with the cut-off, as seen in Table I. Actually, its
value at a large $\mu$ , e.g, at $2\, \rm GeV$ , should be smaller than
the values we give in Table I, since the form factors $h(q^{2})$
should be further suppressed at large $q^{2}$ by perturbative QCD
effects such that the cross sections for the production of $0^{-}1^{-}$
 pair in $e^{+} e^{-}$ collisions (e.g, $e^{+} e^{-} \to \pi\rho$)
will not grow too fast
with energy. This suppression also guarantees  the convergence of the
$Q^{2}$ integral for the Cottingham formula \cite{harari} and make our
results insensitive to the value of the cut-off $\mu$.

In conclusion, we can say that the small mass difference
 $B^{+}-B^{0}$ can be understood, in the $m_{Q}\to\infty$ limit, as a sum
 of two contributions of opposite sign and similar size that remain
 finite in this limit. The electromagnetic contribution has been
 computed by the Cottingham formula and has a small dependence on the
 renormalization mass scale $\mu$. The HQET results are very similar
to those obtained at finite $b$ mass.  In the case of $D^{+}-D^{0}$ the
 contributions have the same sign and  add up; in this case
numerical results show
 deviations of $\simeq 30\%$
as compared to the predictions obtained in the HQET limit.

\vspace{2cm}
\noindent{\bf Acknowledgments\\}
\noindent We thank N. Paver for interesting discussions.

\newpage
\begin{center}
  \begin{Large}
  \begin{bf}
  Table Captions
  \end{bf}
  \end{Large}
\end{center}
\vspace{1cm}

\noindent {\bf Table I}\\
\noindent 
Different contributions to the electromagnetic
mass differences in the $B$ and $D$ systems (units are MeV). The first value is
obtained using $\mu = 1 \,\rm GeV$ ; the second value (in parentheses) using
$\mu =  2 \,\rm GeV$.

\vspace{0.5cm}
\noindent {\bf Table II}\\
\noindent 
Electromagnetic and quark-mass contributions to the 
mass differences in the $B$ and $D$ systems (units are MeV)
compared to the experimental
data \cite{PDG}. The e.m. value is an average between the results obtained with
$\mu =  1\; \rm GeV$ and $\mu=2\; \rm GeV$.

\newpage
\begin{table}
\begin{center}
\begin{tabular}{l c c c c } 
& & {\bf Table I} & &  \\  & & & & \\
 \hline \hline
$\delta m $ & Born & $ 1^-$ & 
$1^+$ & total 
\\ \hline \\
$\delta m(D^+ -D^0)$ & $1.72~(2.28) $ & $-0.09~(- 0.34)$ & $
0.004~(-0.007)$ & $1.63~(1.95)$ \\ \\ 
\hline \\
$\delta m(B^+ -B^0)$ & $1.50~(1.87) $ & $-0.18~(-0.41)$ & $0.005~(0.01)$ & 
$1.33~(1.47)$ \\ \\ 
\hline \hline
\end{tabular}
\end{center}
\end{table}

\begin{table}
\begin{center}
\begin{tabular}{l c c c c } 
& & {\bf Table II} & &  \\  & & & & \\
 \hline \hline
$\delta m $ & e.m. & quark mass & 
 total& exp. \cite{PDG} 
\\ \hline \\
$\delta m(D^+ -D^0)$ & $+1.79\pm 0.16 $ & $+2.54\pm 0.21$ & $+4.33\pm 0.37$ & 
$+4.78\pm 0.10$ \\ \\ 
\hline \\
$\delta m(B^+ -B^0)$ & $+1.40\pm 0.07 $ & $-2.23\pm 0.27$ & $-0.83\pm 0.34 $ 
& $-0.35\pm 0.29$ \\ \\ 
\hline \hline
\end{tabular}
\end{center}
\end{table}

\end{document}